\pdfoutput=1
\documentclass[11pt,twoside,a4paper,cmspaper,final,collab]{cms-tdr}
\def\svnVersion{9aeee30}\def\svnDate{2023/05/18}\def\cmsCernNoTag{CERN-EP-2022-074}\def\cmsCernDate{\today}\def\cmsMessage{Published in Physics Letters B as \href{http://dx.doi.org/10.1016/j.physletb.2022.137534}{\doi{10.1016/j.physletb.2022.137534}.}}
\begin{document}\cmsNoteHeader{HIG-20-008}

\newcommand{\zz}{\PZ\PZ^{*}}
\ifthenelse{\boolean{cms@external}}{\providecommand{\cmsLeft}{upper\xspace}}{\providecommand{\cmsLeft}{left\xspace}}
\ifthenelse{\boolean{cms@external}}{\providecommand{\cmsRight}{lower\xspace}}{\providecommand{\cmsRight}{right\xspace}}
\newlength\cmsTabSpace
\setlength\cmsTabSpace{1.2ex}

\cmsNoteHeader{HIG-20-008}
\title{Search for Higgs boson decays into \texorpdfstring{\PZ}{Z} and \texorpdfstring{\JPsi}{JPsi} and for Higgs and  \texorpdfstring{\PZ}{Z} boson decays into \texorpdfstring{\JPsi}{JPsi} or \texorpdfstring{\PgU}{Upsilon} pairs in \texorpdfstring{$\Pp\Pp$}{pp} collisions at \texorpdfstring{$\sqrt{s} = 13\TeV$}{sqrt(s) = 13 TeV}}

\date{\today}

\abstract{
Decays of the Higgs boson into a \PZ boson and a \JPsi or $\PGyP{2S}$ meson are searched for in four-lepton final states with the CMS detector at the LHC. A data set of proton-proton collisions corresponding to an integrated luminosity of 138\fbinv is used. Using the same data set, decays of the Higgs and \PZ boson into quarkonium pairs are also searched for. An observation of such decays with this sample would indicate the presence of physics beyond the standard model. No evidence for these decays has been observed and upper limits at the 95\% confidence level are placed on the corresponding branching fractions ($\mathcal{B}$). Assuming longitudinal polarization of the Higgs boson decay products, 95\% confidence level observed upper limits for $\mathcal{B}(\PH\to\PZ\JPsi)$ and $\mathcal{B}(\PH\to\PZ\PGyP{2S})$ are $1.9\times 10^{-3}$ and $6.6\times 10^{-3}$, respectively.
}

\hypersetup{
pdfauthor={CMS Collaboration},
pdftitle={Search for Higgs boson decays into Z and J/Psi and for Higgs and Z boson decays into J/Psi or Y pairs in pp collisions at sqrt(s)=13TeV},
pdfsubject={Rare decays},
pdfkeywords={CMS, standard model physics, Higgs boson, Z boson, rare decays}}

\maketitle

\section{Introduction}
A boson with a mass of about 125\GeV was discovered by the ATLAS and CMS Collaborations 
at the CERN LHC in 2012~\cite{Chatrchyan2013,201230,Chatrchyan:2013lba}.
Comprehensive studies in various decay channels and production modes followed, 
and combined measurements show that the properties of the 
new boson are, so far, consistent with the standard model (SM)
predictions for the Higgs boson (\PH) \cite{Aad:2015zhl,CMS:2015kwa,CMS:2018lkl}.
Rare exclusive decays of the \PH to mesons provide experimentally clean 
final states to study deviations of Yukawa couplings from SM predictions that 
cannot be obtained with inclusive measurements.
Several models beyond the SM predict enhanced Yukawa couplings to fermions~\cite{LHC-YR4}, 
leading to branching fractions ($\mathcal{B}$) that are enhanced by up to three orders of magnitude.
Examples include the Giudice--Lebedev model of quark masses~\cite{giudice},
the two Higgs doublet model~\cite{Egana-Ugrinovic:2019dqu},
the single Higgs doublet model with the Froggatt--Nielsen
mechanism~\cite{froggatt}, and Randall--Sundrum
models~\cite{randall,Huber:2000ie}.
The required sensitivity for observing Yukawa couplings
to the second- and first-generation fermions has not yet been
reached, although recently the CMS Collaboration published evidence
for the Higgs boson decay to a pair of muons~\cite{hmumu,2021135980}.
The observed upper limit for $\mathcal{B}(\PH\to\ccbar)$ times the 
cross section for \PH production in association with vector bosons
as measured by the ATLAS and CMS experiments is found to be 26 and
14 times the SM expectation~\cite{ATLASccbar,CMSccbar}, respectively.

\begin{figure*}[htb!]
\centering
\includegraphics[width=0.30\textwidth]{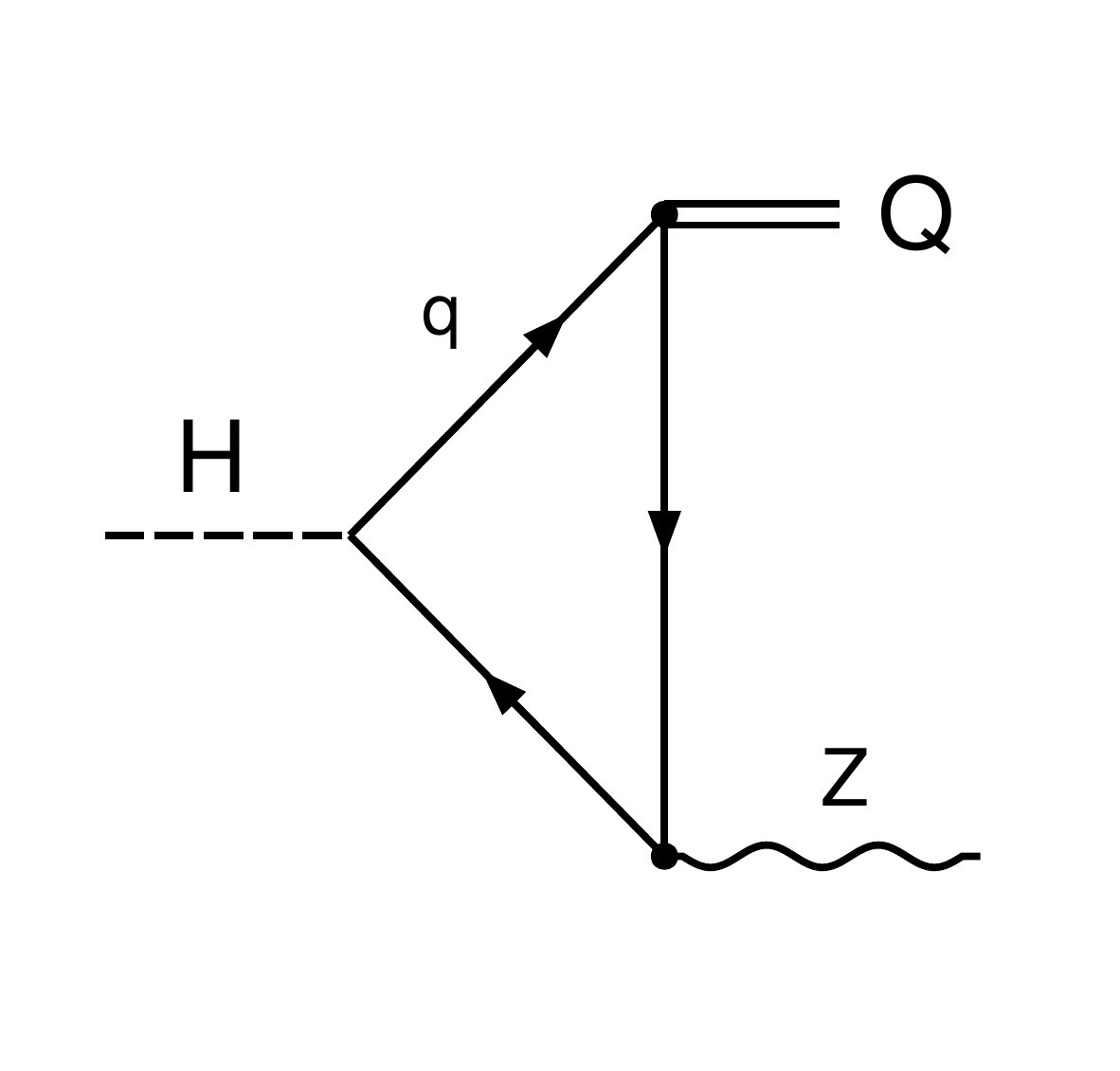}
\includegraphics[width=0.30\textwidth]{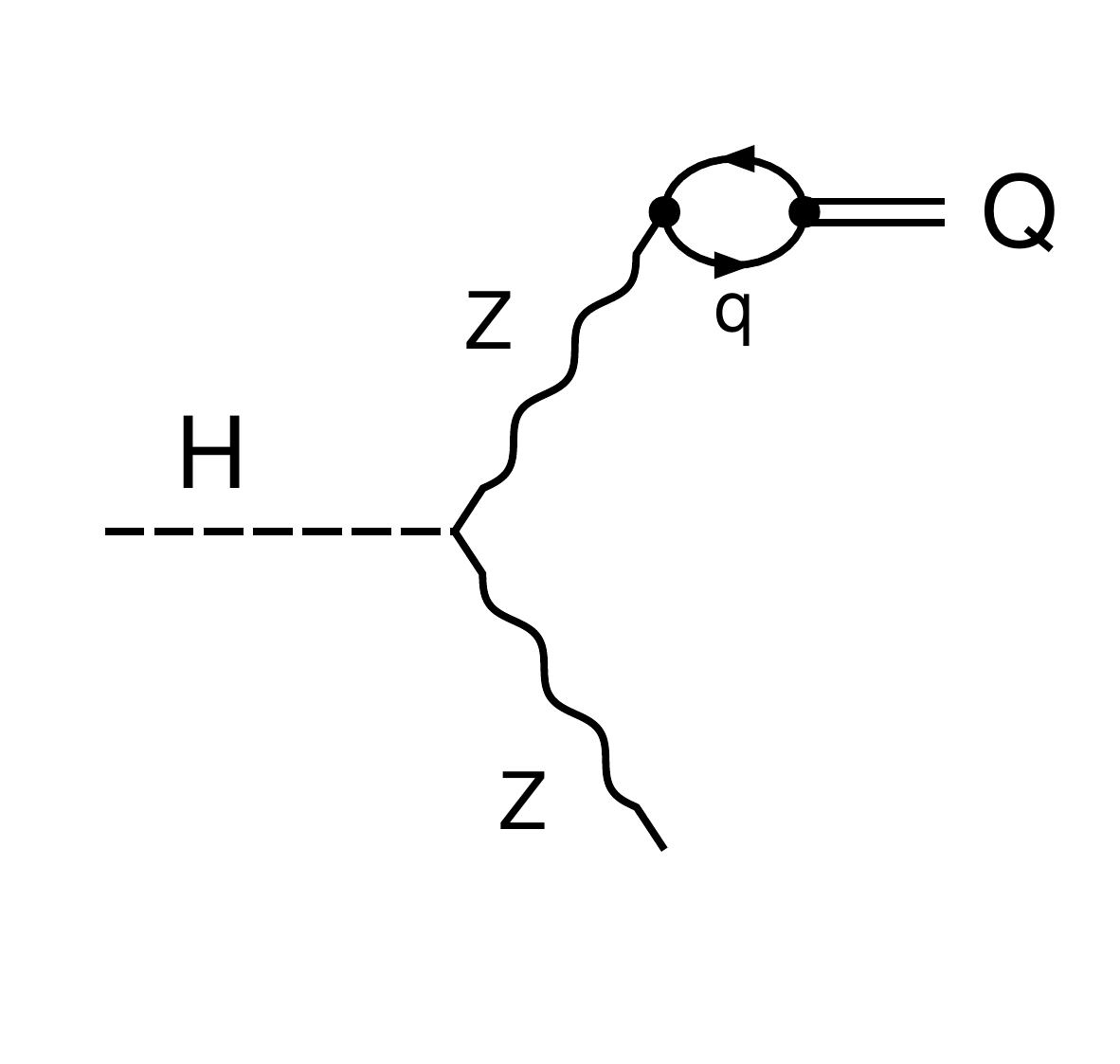}
\includegraphics[width=0.30\textwidth]{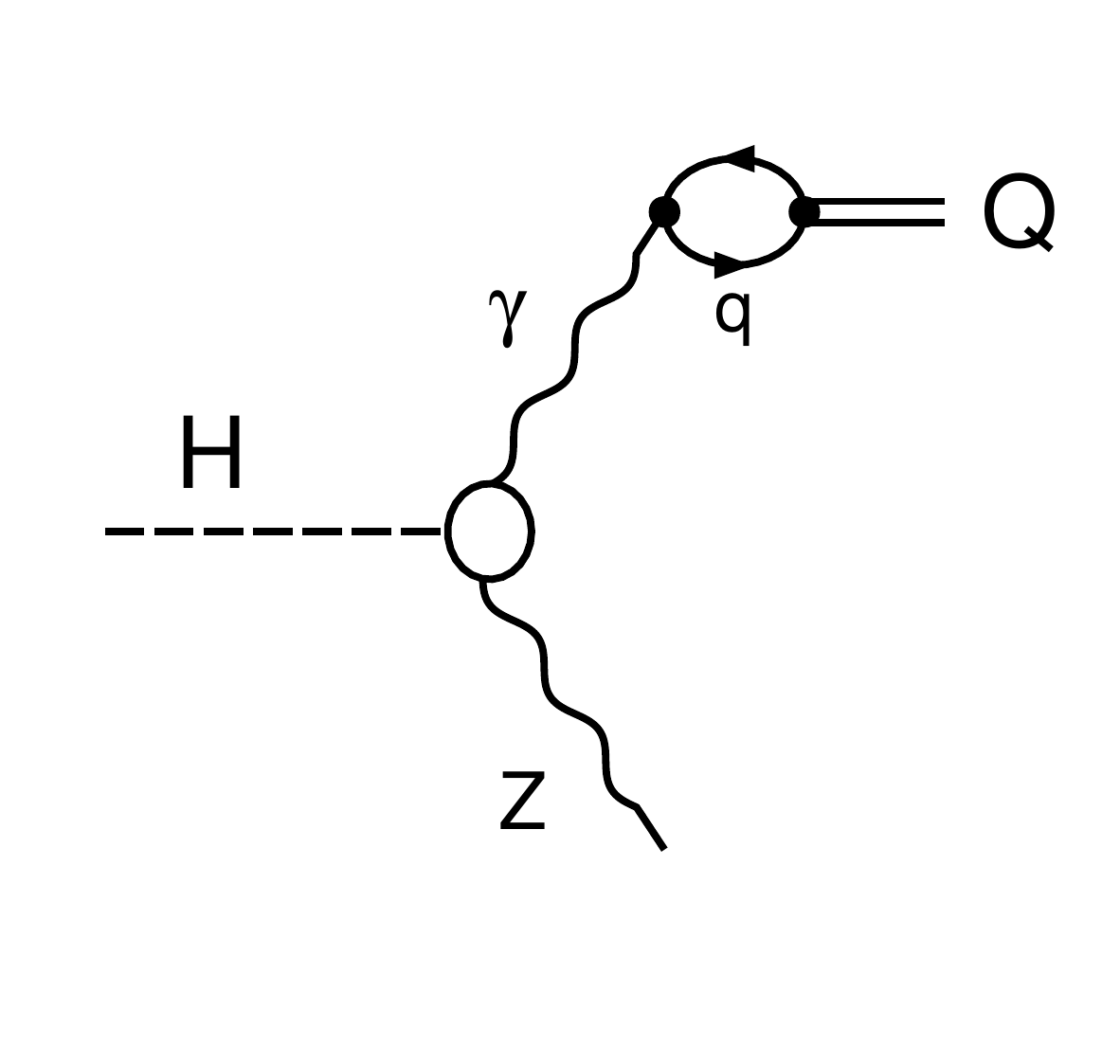}
\caption{Sample Feynman diagrams depicting direct (left) and indirect (middle, right)
quark coupling contributions to the $\PH\to\PZ\mathrm{Q}$ decay,
where $\mathrm{Q}$ represents a quarkonium state and $\mathrm{q}$ is a quark, which
in the leftmost diagram is either of charm or bottom flavor.
The diagrams represent Higgs boson decays into quarkonium pairs when
replacing the bottom section with the upper half in each.
\label{fig:feynman}}
\end{figure*}

The first class of processes that is considered here is
the decay of the Higgs boson into a \PZ boson and
a vector meson quarkonium state ($\mathrm{Q}$)~\cite{Alte:2016yuw,LHC-YR4,2014PhLB_7}.
The relevant SM Feynman diagrams for the
decays $\PH\to\PZ\mathrm{Q}$ are shown in Fig.~\ref{fig:feynman}.
The first diagram in Fig.~\ref{fig:feynman} represents
contributing amplitudes at leading order (LO), where the
Higgs boson directly couples to a quark-antiquark
pair that radiates a \PZ boson and forms the meson.
The last two diagrams
depict indirect contributions to the decay amplitude.
The last diagram corresponds to one-loop amplitudes as
indicated by the blob, as well as tree-level effective vertices~\cite{Alte:2016yuw}.
The branching fractions of these processes in the SM are
$\mathcal{B}(\PH\to\PZ\JPsi) = 2.3 \times 10^{-6}$~\cite{Alte:2016yuw,2014PhLB_7,gao}
and $\mathcal{B}(\PH\to\PZ\PGyP{2S}) = 1.7 \times 10^{-6}$~\cite{gao}.
The main source of background events in these final
states is from production of a \PZ boson in association
with a genuine~\cite{associatedproduction} or misidentified
meson candidate.
New physics could affect the direct boson couplings or
could enter through loops and alter the interference
pattern between the amplitudes. Any of those possibilities
can enhance branching fractions with respect to the SM
predictions.
For the rare decays $\PH\to\PZ\JPsi$ and
$\PH\to\PZ\PGyP{2S}$, the $\PH\to\PZ\PGg^{\ast}$ amplitude
contributes significantly. Hence, these rare decays provide
complementary information to the decay $\PH\to\PZ\PGg$,
both in and beyond the SM~\cite{gao,Alte:2016yuw}.
The decays of the Higgs boson into $\PZ\JPsi$ and $\PZ\PGhc$
have been searched for by the ATLAS Collaboration in hadronic final states,
reaching 95\% confidence level (\CL) upper limits on the branching fraction
of the Higgs boson that exceed 100\%~\cite{Aad:2020hzm}.
Recently, the CMS Collaboration published upper limits
on the branching fraction for $\PH\to\PZ\PGr$ and $\PH\to\PZ\PGf$
at 95\% \CL that are larger than the expected SM branching fraction by
more than a factor of 700~\cite{Sirunyan:2020mds}.
 
A second related class of processes is the Higgs boson decay into pairs of quarkonia.
The Feynman diagrams are variants of the ones depicted in
Fig.~\ref{fig:feynman}: in each diagram, the on-shell \PZ boson
in the lower part is replaced by a quarkonium decay, similar
to the process depicted in the upper part.
In the rightmost diagram, both vector bosons could be also gluons,
in which case additional soft-gluon exchange occurs.
The importance of the measurement of such decays has been
pointed out in Refs.~\cite{soni1979,keung1983,doroshenko1987,kartvili1988}.
Using a phenomenological approach for the direct \PH-$\qqbar$
coupling, Ref.~\cite{soni1979} finds that the dominant
quarkonium pair decay mode is $\PH\to\PGU\PGU$ with an
estimated branching fraction of O($10^{-5}$).
More recently, Ref.~\cite{kartveli} predicts values of 
$\mathcal{B}(\PH\to\JPsi\JPsi) = 1.5 \times 10^{-10}$
and $\mathcal{B}(\PH\to\PGU\PGU) = 2 \times 10^{-9}$
assuming the dominance of indirect amplitudes.
Inclusion of the mechanism where the Higgs boson couples directly
to charm or bottom quarks, which then hadronize to heavy quarkonia,
in the calculation in Ref.~\cite{Bodwin:2013gca} leads to an increase
by an order of magnitude in the related $\mathcal{B}(\PH\to\JPsi\PGg)$.
The Higgs boson is expected to couple to quarkonium pairs
that include radially excited states with comparable
strength~\cite{PhysRevD.100.053002}.
Recently, a first search for the decays
$\PH\to\JPsi\JPsi$ and $\PH\to\PGUP{nS}\PGUP{mS}(\mathrm{n},\mathrm{m}=1,2,3)$
was performed by the CMS Collaboration~\cite{HIG-18-025}.
 
Related to these two classes of processes is the decay of the Higgs boson
into a photon and a vector
meson~\cite{Bodwin:2013gca,Kagan:2014ila, Alte:2016yuw}.
The 95\% \CL upper limits on the branching fractions of the
Higgs boson into $\PGg\JPsi$, $\PGg\PGr$, and
$\PGg\PGf$ are found to be two orders of magnitude larger than their
expected values in the SM~\cite{Sirunyan:2018fmm,Aaboud:2018txb,Aaboud:2017xnb}.
For the $\PGg\PGyP{2S}$ and $\PGg\PGUP{nS}$ decays,
the corresponding upper limits are, respectively, three and five orders
of magnitude larger than the expected SM branching fractions.
 
This Letter presents the first search for decays of the
125\GeV Higgs boson into a \PZ boson and a \JPsi or
$\PGyP{2S}$ meson in four-lepton final states.
The \PZ boson is reconstructed from its decays
into \MM or \EE, the \JPsi meson from its decay into \MM, and the
$\PGyP{2S}$ from its inclusive decay into $\JPsi X$ (feed-down), 
where $X$, which is mostly $\PGp\PGp$, is not reconstructed.
Furthermore, an update of
Higgs boson searches in $\JPsi\JPsi$ and $\PGUP{nS}\PGUP{mS}$ 
decay channels with the full available data sample is presented. 
New channels are accessed 
via the inclusive decay of $\PGyP{2S}$ into a \JPsi meson.
For the \PGUP{nS} states the possibilities that they are
the result of feed-down transitions from higher \PGU states 
before decaying into muon pairs are included.
Finally, this Letter also presents the search for decays of the
\PZ boson into quarkonium pairs.
The SM prediction for $\mathcal{B}(\PZ\to\JPsi\JPsi)$
calculated in the framework of nonrelativistic quantum
chromodynamics (QCD) and
leading twist light cone models is of the order of
$10^{-12}$~\cite{zdecaytheory}.
The \PZ decay into a \JPsi meson and a lepton pair, which is
dominated by the electromagnetic fragmentation process, was
observed at a rate consistent with SM predictions~\cite{Zmumu}.
 
The results presented in this Letter are based on proton-proton
($\Pp\Pp$) collision data recorded in 2016--2018 with
the CMS detector at a center-of-mass energy of
$\sqrt{s}=13\TeV$, amounting to an integrated
luminosity of 138\fbinv in the $\PZ\JPsi$ channel
and 133\fbinv in the quarkonium pair channels,
where the second number is slightly smaller
due to a delayed trigger deployment.

\section{The CMS detector}
The CMS apparatus~\cite{CMS-Jinst} is a multipurpose, nearly hermetic detector, 
designed to trigger on and identify electrons, muons, photons, 
and hadrons. 
The CMS superconducting solenoid provides an axial magnetic field 
of 3.8\unit{T}. Within the solenoid volume are a silicon pixel and strip
tracker, a lead tungstate crystal electromagnetic calorimeter (ECAL),
and a brass and scintillator hadron calorimeter, each composed
of a barrel and two endcap sections. 
Forward calorimeters extend the pseudorapidity ($\eta$) coverage 
provided by the barrel and endcap detectors. Muons are detected in
gas-ionization chambers embedded in the steel flux-return 
yoke outside the solenoid.
A more detailed description of the CMS detector, 
together with a definition of the coordinate system used and the 
relevant kinematic variables, can be found in Ref.~\cite{CMS-Jinst}.

The silicon tracker measures charged particles within the
range $\abs{\eta} < 2.5$. 
It consists of pixel and strip detector modules. 
An entirely new pixel detector has been installed 
during a technical stop between the 2016 and 2017 data-taking periods,
featuring an all-silicon device with four layers in the barrel
and three disks in the endcaps~\cite{newpixel}, providing four
pixel detector measurements within a range 
$\abs{\eta} < 3$.

Muons are measured in the range $\abs{\eta} < 2.4$, with
detection planes made using three technologies: drift tubes,
cathode strip chambers, and resistive-plate chambers.
The single-muon trigger efficiency exceeds 
90\% over the full $\eta$ range, and the efficiency to
reconstruct and identify muons is greater than 96\%. 
Matching muons to tracks measured in the silicon tracker
results in relative transverse momentum (\pt) resolutions
of 1\% in the barrel and 3\% in the endcaps
for muons with \pt up to 100\GeV, and better than
7\% in the barrel for muons with \pt up to 1\TeV~\cite{Sirunyan:2018}. 

An electron is reconstructed by combining an energy
measurement in the ECAL with a momentum 
measurement in the tracker. 
The ECAL consists of 75,848 lead tungstate crystals,
which provide coverage of $\abs{\eta} < 1.48$ in the barrel
region and $1.48 < \abs{\eta} < 3.00$ in the two endcap 
regions (EE). Preshower detectors consisting of two planes 
of silicon sensors interleaved with a total of 
three radiation lengths of 
lead are located in front of each EE detector.
The momentum resolution for electrons with
\mbox{$\pt \approx 45\GeV$} from $\PZ \to \EE$ decays 
ranges from 2 to 5\%. It is generally better in the barrel
region than in the endcaps, and also depends on the 
bremsstrahlung energy emitted by the electron as it traverses
the material in front of the ECAL~\cite{Khachatryan:2015hwa}.

The candidate vertex with the largest value of summed 
physics-object $\pt^2$ is taken to be the primary 
$\Pp\Pp$ interaction vertex. The physics objects 
are the jets, clustered using the jet finding 
algorithm~\cite{Cacciari:2008gp,Cacciari:2011ma} with 
the tracks assigned to candidate vertices as inputs, 
and the associated missing transverse momentum, taken 
as the negative vector \pt sum of those jets. 
Other collision vertices in the event are considered to 
have originated from additional inelastic $\Pp\Pp$ 
collisions in each bunch crossing, referred to as pileup 
(PU). The average number of PU interactions during the 
2016 data-taking period was 23, and increased to 32 during 
the 2017 and 2018 data-taking periods.

Events of interest are selected using a two-tiered trigger system. 
The first level, composed of custom hardware processors, 
uses information from the calorimeters and muon detectors to 
select events at a rate of around 100\unit{kHz} within a fixed 
latency of about 4\mus~\cite{Sirunyan:2020zal}. The second 
level, known as the high-level trigger, consists of 
a farm of processors running a version of the full event 
reconstruction software optimized for fast processing, 
and reduces the event rate to around 1\unit{kHz} before 
data storage~\cite{Khachatryan:2016bia}.

Dedicated triggers were deployed to enhance the selection of
events of interest for the present study. 
Single-lepton (muon or electron) triggers are used 
to study the $\PZ\JPsi$ channel. The single muon trigger requires
an isolated muon with $\pt > 27\GeV$. The single
electron trigger requires an isolated electron 
having $\pt > 27\,(35, 32)\GeV$ during the year 
2016 (2017, 2018). 
The meson-specific triggers are used to study quarkonium 
channels. They require the presence of at least
three muons with $\pt > 2\GeV$. Two of those must be
oppositely charged and originate from a common vertex with a
probability greater than 0.5\%, as determined by a Kalman
vertex filter~\cite{kalman}, thus suppressing random combinations 
of two muons. The \JPsi- (\PGU-) specific trigger
requires a dimuon system invariant mass to be between 2.95
and 3.25 (8.5 and 11.4)\GeV. The dimuon system 
\pt is required to exceed 3.5\GeV for the years 2017 and 2018. 

\section{Simulated samples}
Simulated samples of the \PH and \PZ boson signals are 
used to estimate the expected signal yields and model the
distribution of signal events in the four-lepton
invariant mass. 
The SM Higgs boson signals are simulated
at next-to-leading order (NLO) in perturbative QCD 
with the \POWHEG v2.0 Monte Carlo event
generator~\cite{Alioli:2008tz,Nason:2009ai}, which includes
the gluon-gluon fusion and vector boson fusion production 
processes. The parton distribution function (PDF) set used is 
NNPDF3.1~\cite{Ball:2017nwa}.
The JHUGen 7.1.4 generator~\cite{jhugenerator1,jhugenerator2} 
is used to decay the Higgs boson into 
\PZ bosons and $\mathrm{Q}$ mesons.
The generator is interfaced with \PYTHIA 8.226~\cite{Sjostrand:2014zea}
for parton showering, hadronization, and underlying event simulation
using the CUETP8M1~\cite{Khachatryan:2015pea} tune.
The total SM Higgs boson production cross section for the
calculation of branching fractions is taken from the 
LHC Higgs cross section working group~\cite{LHC-YR4}.

The \PZ bosons are  
produced and decayed with the \PYTHIA 8.226 generator~\cite{Sjostrand:2014zea}
which implements the LO matrix element calculation interfaced 
with parton showering, hadronization, and underlying event simulation
for which the tune CUETP8M1~\cite{Khachatryan:2015pea} is used.
For the calculation of the branching fraction, 
the inclusive SM \PZ boson production cross section 
is obtained according to Ref.~\cite{SMP-15-004}, where the prediction 
includes the next-to-next-to-leading order (NNLO) QCD contributions, and
the NLO electroweak corrections from \FEWZ 3.1~\cite{fewz} calculated using 
the NLO PDF set NNPDF3.0. The generated events are then reweighted to match 
the \pt-spectrum of the \PZ boson predicted at 
NLO~\cite{Alioli:2008tz,Nason:2009ai,Alwall:2014hca}. 

The generated events are processed through a detailed simulation 
of the CMS detector based on \GEANTfour~\cite{GEANT4}. 
Simulated events include additional $\Pp\Pp$ interactions.
Events are then reweighted to match the PU profile observed in data.

The acceptance of the final states changes with the angular
distribution of leptons in the decay. 
The distribution of the decay angle $\theta$, defined as the
angle between the positive lepton momentum in the
rest frame of the intermediate 
particle (\JPsi meson or \PGU meson or \PZ boson)
with respect to the direction of this intermediate 
particle in the rest frame of the
parent particle (\PH or \PZ boson), is proportional to 
$(1 + \lambda_\theta \cos^2\theta)$, where $\lambda_\theta$
is the average polar anisotropy parameter~\cite{faccioli}.
According to Refs.~\cite{kartveli,zdecaytheory,PhysRevD.98.115037}, 
longitudinal polarization is expected for the \PZ boson 
and assumed for mesons.
In this Letter, the nominal results are obtained using a signal
acceptance calculated for the longitudinally polarized case
$(\lambda_\theta = -1)$. 
Two other extreme scenarios where the intermediate particles
are both either fully transversely polarized 
($\lambda_\theta = +1$) or unpolarized ($\lambda_\theta = 0$)
are also considered. 
No azimuthal anisotropies are considered. 

\section{Event reconstruction}
In a first step, events with at least $2\PGm$ and two additional 
$\Pell$ ($\Pell = \Pe$ or \PGm) are selected. The \JPsi and \PGU 
candidates are built from \MM pairs, and \PZ candidates from 
\MM or \EE pairs.

Muons are reconstructed by combining information from the
silicon tracker and the muon system~\cite{Sirunyan:2018}.
The matching between tracks reconstructed in each of the
subsystems proceeds either outside-in, starting from a track 
in the muon system, or inside-out, starting from a track 
provided by the silicon tracker. In the latter case, 
tracks that match track segments in only one or two 
detector layers
of the muon system are also considered in 
the analysis to collect very low \pt muons that may 
not have sufficient energy to penetrate the
entire muon system. The muons are selected from the
reconstructed muon track candidates that match with 
at least one segment in any muon detector layer.
The number of silicon tracker layers with 
hits used in the muon track candidate has to be greater 
than 5 and include at least one pixel detector layer.
To suppress muons originating from non-prompt hadron decays, 
the impact parameter of each muon track, computed with 
respect to the position of the primary $\Pp\Pp$ interaction 
vertex, is required to be less than 0.3\,(20.0)\unit{cm} 
in the transverse plane (along the longitudinal axis).
Events with at least four such muons with $\pt>3\GeV$ and 
$\abs{\eta} <2.4$ and zero sum of charges are accepted. 
To measure the isolation of the leading (highest \pt) 
muon candidate from other hadronic activity in the event, a 
cone of size ${\Delta}R=\sqrt{\smash[b]{(\Delta\eta)^{2}+(\Delta\phi)^{2}}} = 0.3$ 
is constructed around its momentum direction, where $\phi$ is 
the azimuthal angle in radians. The \pt sum of the reconstructed 
inner-detector tracks originating from the primary $\Pp\Pp$ interaction 
vertex within the cone has to be less than 50\% of the muon's \pt.
The leading muon \pt is excluded from the sum
and the subleading muon \pt is also excluded 
if this muon falls within
the isolation cone of the leading muon.

The energy of electron candidates is determined from a 
combination of the track momentum at the main interaction 
vertex, the corresponding ECAL cluster energy, and the 
energy sum of all bremsstrahlung photons attached to 
the track~\cite{Khachatryan:2015hwa}. The dielectron 
mass resolution for $\PZ\to\EE$ decays when both 
electrons are in the ECAL barrel is 1.9\%, and is 
2.9\% when they are in the endcaps. 
To reduce contamination from particles incorrectly 
identified as electrons, reconstructed electrons are 
required to pass a multivariate electron identification 
discriminant. This discriminant, which is described 
in Ref.~\cite{Khachatryan:2015hwa}, combines information 
about the quality of tracks, the shower shape, kinematic 
quantities, and hadronic activity in the vicinity of the 
reconstructed electron. Isolation variables are also 
included among the discriminant's inputs. Therefore, no 
additional isolation requirements are applied. The 
selection based on the multivariate identification 
discriminant has an electron identification efficiency of 
90\% while the rate of misidentifying other particles as 
electrons is 2--5\%. 
Events with at least two oppositely charged electrons  
with $\pt>3\GeV$ and $\abs{\eta} <2.5$ and two oppositely
charged muons are accepted.

Further selection criteria are applied to the different four-lepton
final states to achieve the lowest expected upper limit at 
95\% \CL~\cite{junk,read,combiner}. 
This optimization is performed with data with the 
signal regions removed~\cite{blinding} and replaced by 
simulated events, and with the simulated signal shape.

\begin{figure}[htb!]
  \centering
    \includegraphics[width=0.49\textwidth]{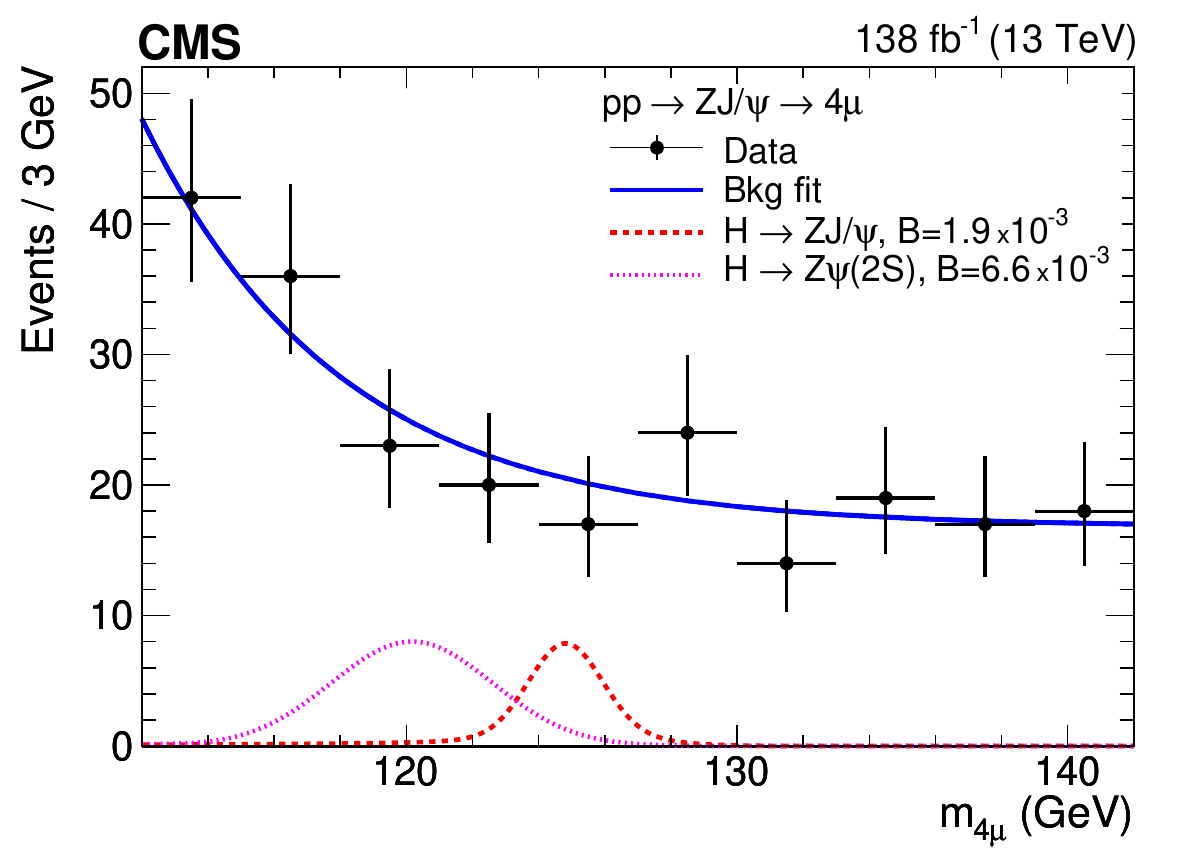}
    \includegraphics[width=0.49\textwidth]{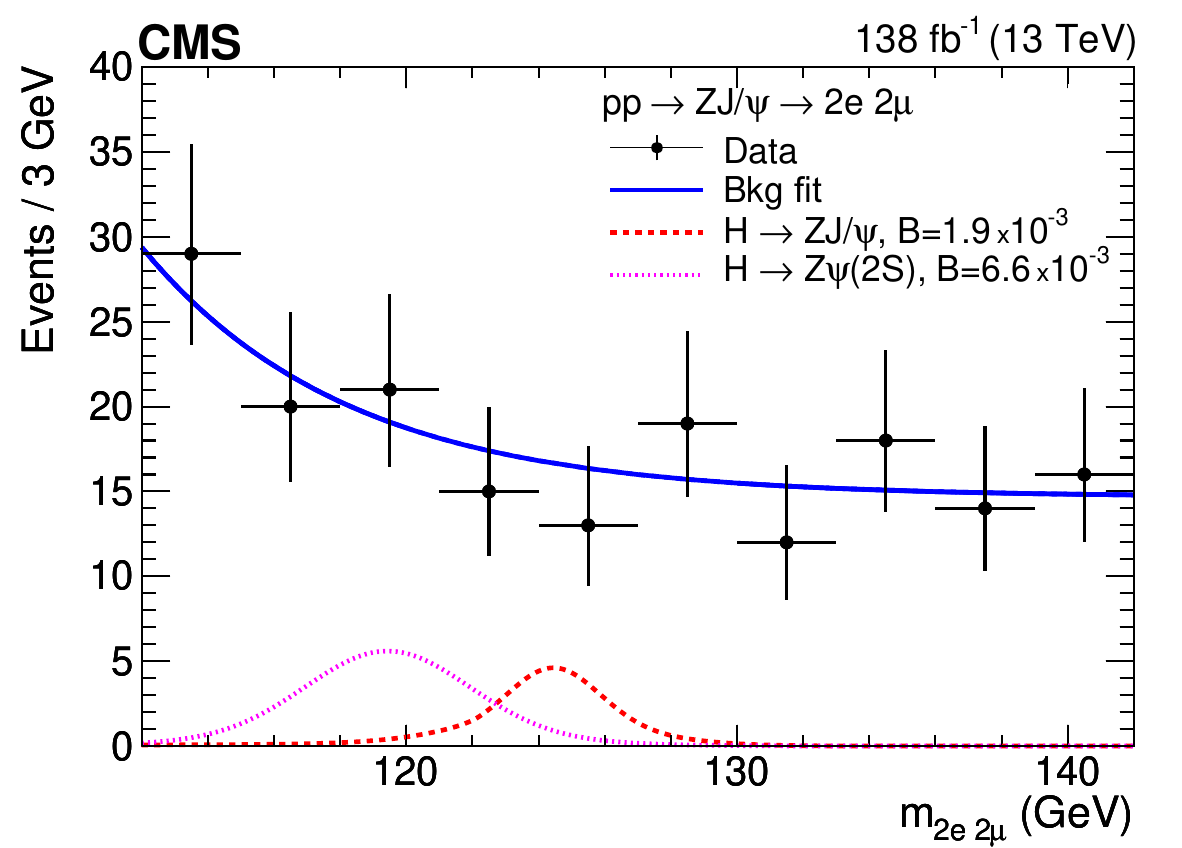}
    \caption{\label{fig:kineplot1ZJ} Four-lepton invariant mass distributions, 
      (\cmsLeft) for $\PZ\JPsi\to4\Pgm$ candidates and (\cmsRight) for   
       $\PZ\JPsi\to2\Pe2\Pgm$ candidates.
      The result of the maximum likelihood fit to the background (Bkg)
      is superimposed (solid blue line). For illustrative purposes, the plots show 
      simulated $\PH\to\PZ\JPsi$ (dashed red line) and 
      $\PH\to\PZ\PGyP{2S}$ (dotted magenta line) signals 
      normalized to their observed upper limit branching fractions 
      at 95\% \CL as obtained in this analysis, 
      where \PGyP{2S} decays into $\JPsi + X$.}
\end{figure}

The selection in the $\PZ\JPsi$ channel requires each
dilepton resonance to have $\pt > 5\GeV$, and
the candidate invariant mass to lie in the region 80--100\GeV 
for the \PZ boson (3.0--3.2\GeV for the \JPsi meson). 
The dimuon mass resolution is about 1\%.
Each dilepton must fit to
a common vertex with probability greater than 1\%,
determined by a vertex fit probability. 
The four-lepton candidate must have $\pt > 5\GeV$,
and the fit to a common four-lepton vertex must have a probability 
of greater than 1\%. A total of 230 (177) candidate 
events are found in the 4$\Pgm$ ($2\Pe2\Pgm$) invariant 
mass window of 112--142\GeV. The lower limit of the 
four-lepton invariant mass ($m_{4\Pell}$) range is chosen to exclude the 
region close to the $\PZ\JPsi$ threshold
where $m_{4\Pell}$ changes rapidly, which is difficult to model.
The selection criteria for the decay $\PH\to\PZ\PGyP{2S}$, where
\PGyP{2S} decays inclusively into \JPsi, are identical.
The respective $m_{4\Pell}$ distributions
are shown in Fig.~\ref{fig:kineplot1ZJ}. Non-resonant background
in the dilepton invariant mass distributions is found to contribute
about 20\%. These events are an additional source of background 
which does not concentrate in a specific region of $m_{4\Pell}$, 
and thus is accounted for by an empirical parameterization of the 
background in the fit as described in the next section.
 
\begin{figure}[htb!]
  \centering
    \includegraphics[width=0.49\textwidth]{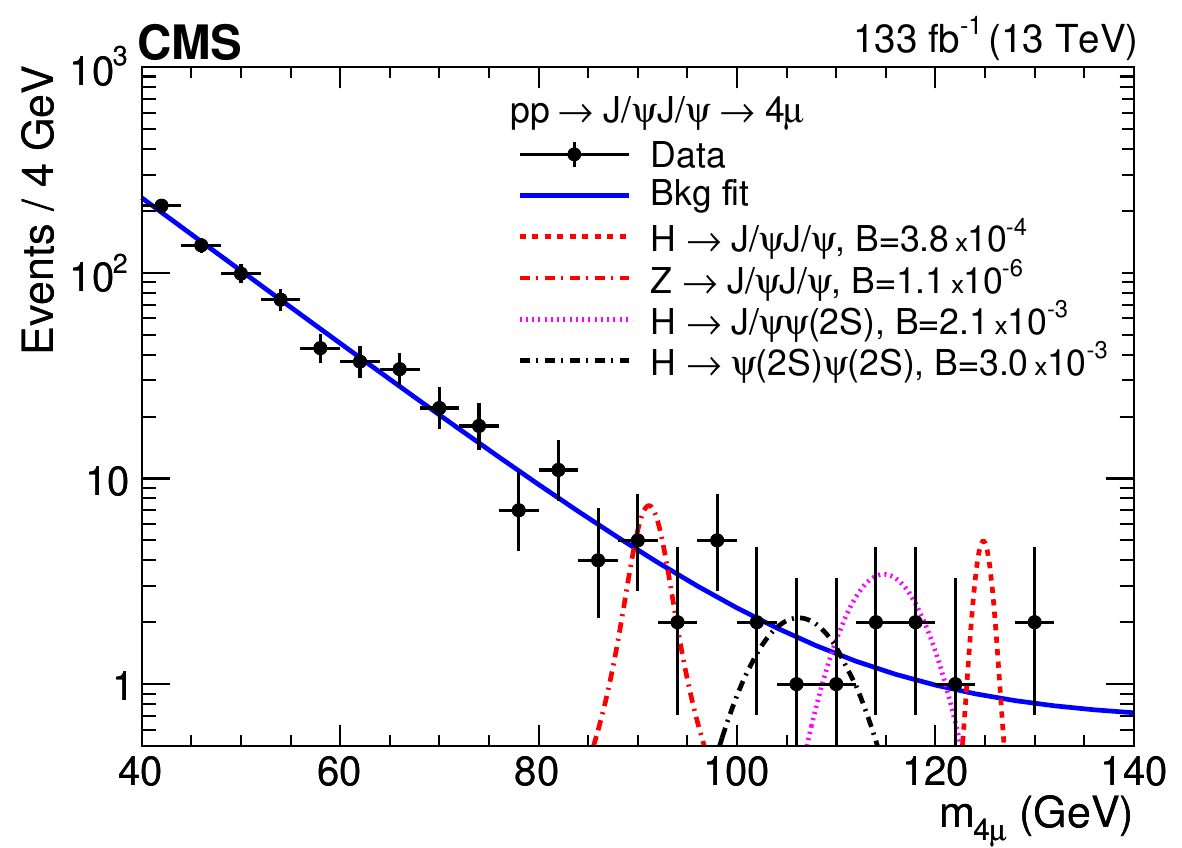}
    \caption{\label{fig:kineplot1JJ}The four-muon invariant mass distribution of 
      $\JPsi\JPsi$ candidates (error bars for empty bins are not shown).
      The result of the maximum likelihood fit to the background (Bkg) 
      is superimposed (solid blue line). For illustrative purposes, the plots 
      show the distributions for simulated signals of the Higgs and $\PZ$ boson 
      decaying into $\JPsi\JPsi$ (dashed and dashed-dotted red lines). 
      The signals for the Higgs boson decays 
      $\PH\to\PGyP{2S}\JPsi$  (dotted magenta line) and 
      $\PH\to\PGyP{2S}\PGyP{2S}$ (dashed-dotted black line) 
      are also shown, where \PGyP{2S} decays into \JPsi. 
      Each signal is normalized to their observed 95\% \CL upper limit branching 
      fraction from this analysis.}
\end{figure}

In the case of the \JPsi pair channel, each dimuon has to be fit to a common vertex 
with a probability greater than 0.5\%.
In addition, the  \JPsi 
candidate must have $\pt > 3.5\GeV$, matching the trigger requirement, 
and the invariant masses of the higher and lower-\pt \JPsi candidates have to be 
within 0.10 and 0.15\GeV, respectively, of the nominal mass of the  \JPsi meson. 
The mass window of the subleading 
\JPsi is wider to allow monitoring the reduction of the sideband population 
as the selection progresses.
In order to suppress contributions from non-prompt hadrons, separately 
produced \JPsi mesons and muons from other sources, the four-muon vertex 
fit probability of \JPsi pairs must be greater than 5\%.
Finally, the absolute value of the difference in rapidity 
($\abs{\Delta Y}$) 
between the two \JPsi candidates has to be less than 3. This criterion 
marginally affects the
signal, while removing about 20\% of  the background events.
After the selection, 720 events are found in data in the 40--140\GeV 
four-muon invariant mass ($m_{4\Pgm}$) range. 
Non-resonant background in the dilepton invariant mass distributions
is found to be negligible. Figure~\ref{fig:kineplot1JJ} shows the $m_{4\Pgm}$ 
distribution of the $\JPsi\JPsi$ candidates.

\begin{figure}[htb!]
  \centering
    \includegraphics[width=0.49\textwidth]{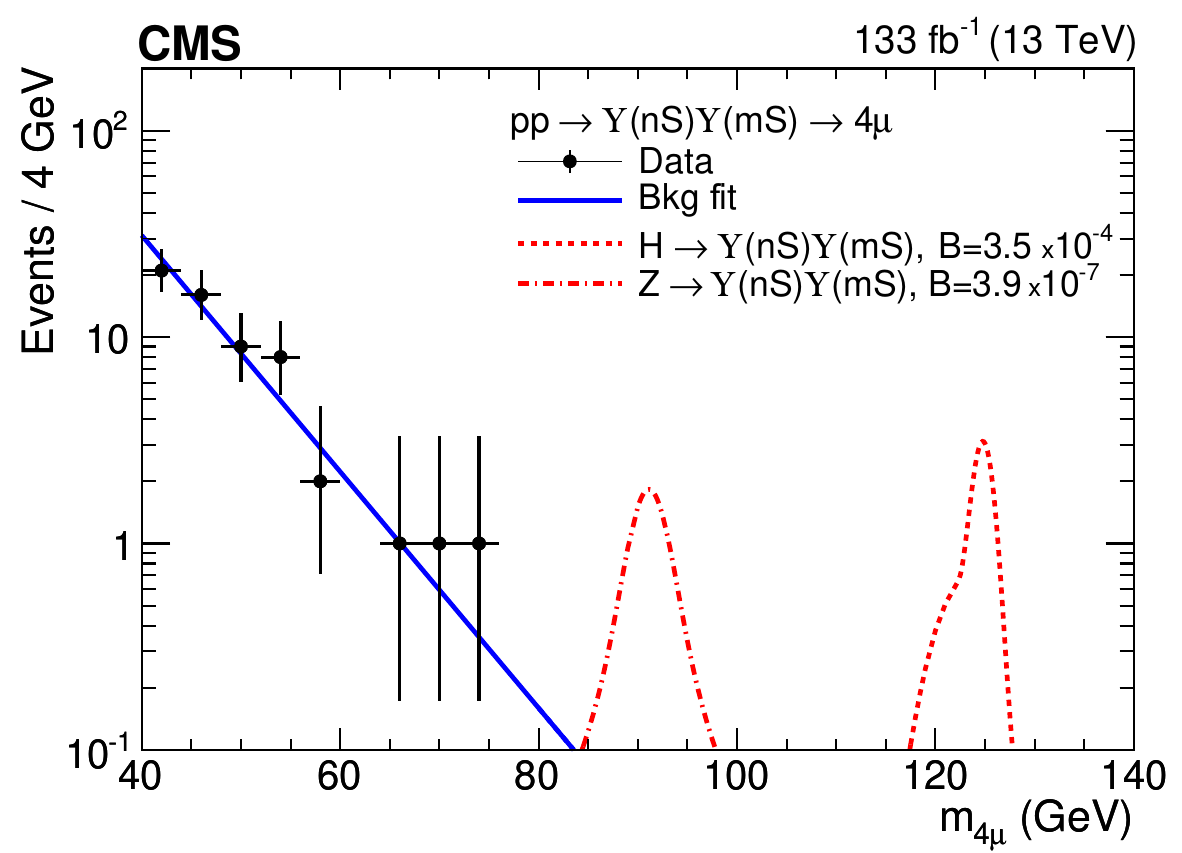}
    \includegraphics[width=0.49\textwidth]{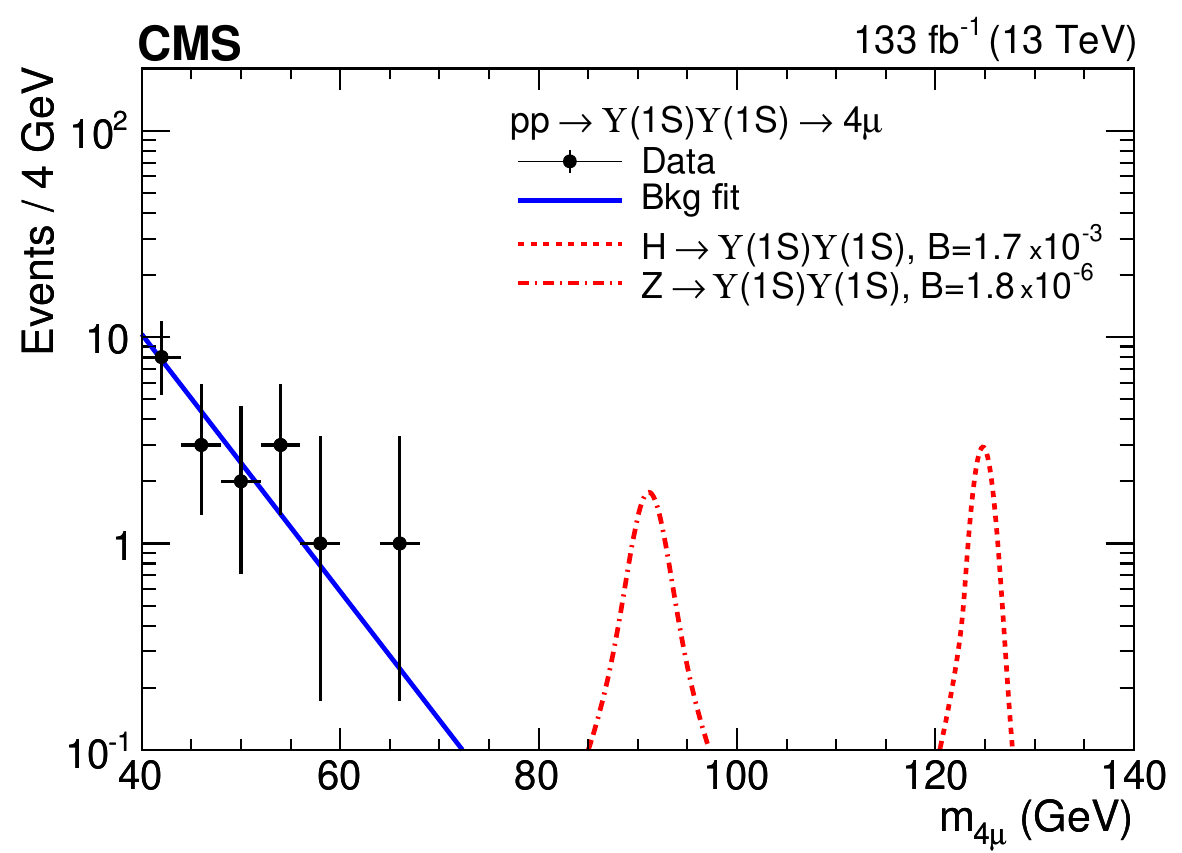}
    \caption{The four-muon invariant mass distributions for 
      $\PGUP{nS}\PGUP{mS}$ (\cmsLeft) and $\PGUP{1S}\PGUP{1S}$ (\cmsRight) 
      candidates (error bars for empty bins are not shown).
      The result of the maximum likelihood fit to background (Bkg) is superimposed
      (solid blue line). For illustrative purposes, the plots show the distributions
      for simulated Higgs and \PZ boson signals (dashed and dashed-dotted red lines) 
      normalized to their observed 95\% \CL upper limit branching fractions
      from this analysis. 
      \label{fig:kineplot1YY}}
\end{figure}

An \PGU pair candidate event must have at least four muons each with \pt $> 4\GeV$.
The $\PGUP{nS} (\mathrm{n} = 1,2,3)$ and \PGUP{1S} candidates are
formed with oppositely charged muon pairs with $\pt > 5\GeV$, and the 
dimuon invariant mass within the range 9.0--10.7\GeV and 9.0--9.7\GeV,
respectively. 
In order to suppress random combinations, dimuon and four-muon objects are required 
to have a vertex fit probability greater than 1\%. 
Between two candidate dimuons, the $\abs{\Delta Y}$ value
has to be less than 2.3 and the azimuthal angle difference has to be greater than
1 radian. The four-muon combination must have $\pt > 5\GeV$ and 
an absolute rapidity of less than 1.7. After applying the selection criteria
in data, 59 $\PGUP{nS}\PGUP{mS}$ (18 $\PGUP{1S}\PGUP{1S}$) candidate 
events are found in the 40--140\GeV $m_{4\PGm}$ range. 
Figure~\ref{fig:kineplot1YY} \cmsLeft and \cmsRight show the 
$m_{4\PGm}$ distributions for $\PGUP{nS}\PGUP{mS}$ and
$\PGUP{1S}\PGUP{1S}$ candidates, respectively.

The differences in the efficiencies between data and simulation for the trigger,
offline lepton reconstruction, identification, and isolation requirements are 
corrected by
reweighting the simulated events with data-to-simulation correction factors,
which are obtained with the ``tag-and-probe'' method~\cite{cite:tagandprobe}
using $\JPsi\to\MM$ and $\PZ\to\Pell^{+}\Pell^{-}$ events. 
The total signal efficiency, including kinematic acceptance, trigger, 
reconstruction, identification, and isolation efficiencies, for the 
$\PZ\JPsi\to4\Pgm$ ($\PZ\JPsi\to2\Pe2\Pgm$) decays with 
longitudinally polarized \JPsi and  \PZ, is found from simulation
to be around 30 (24)\%. For the Higgs boson decays into $\PGUP{nS}\PGUP{mS}$ 
and  $\JPsi\JPsi$, the corresponding total efficiencies
are about 31 and 30\%, respectively. For the \PZ boson, the corresponding 
values are about 28 and 32\%.

\section{Signal extraction and systematic uncertainties}
\label{sec:fit}
Unbinned extended maximum likelihood fits~\cite{roofit} to the 
$m_{4\Pell}$ distributions are performed. 
Yields for signals and backgrounds are free parameters in the fit.
The background shapes in the $m_{4\Pell}$ 
distributions are obtained from data and are described 
by an exponential plus constant function. 
For the $\PZ\JPsi\to4\Pgm$ channel, the \PH signal is 
parameterized with a sum of a Gaussian and a Crystal 
Ball function~\cite{crystalball}, 
and for the $\PZ\JPsi\to2\Pe2\Pgm$ channel with two Crystal Ball functions. 
Similarly, the \PH signal in the \JPsi 
pair channel is described with a double-Gaussian function,
and in the \PGU pair channel with a combination of a Gaussian 
and Crystal Ball functions. In all combined functions the mean 
is a common parameter.
The simulated \PZ boson signal is described with 
a Voigtian function with the resonance width fixed to
the world-average value~\cite{pdg}. The mass resolution 
and the mean are taken from the fit to the simulation,
and they are fixed in the fit to data.
The background function from the fit to
data in the $\PZ\JPsi$ and $\JPsi\JPsi$ channels
are superimposed as solid blue lines in 
Figs.~\ref{fig:kineplot1ZJ} and \ref{fig:kineplot1JJ}, 
respectively. In the \PGU pair sample, no events are 
observed above the $m_{4\Pgm}$ of 80\GeV.
The $m_{4\Pgm}$ distribution below 80\GeV 
is well described solely by an exponential function. 
Figure~\ref{fig:kineplot1YY} shows the observed $m_{4\Pgm}$ 
distribution with the fit superimposed.

Separate fits are performed to the $m_{4\Pell}$
distributions for the different signal hypotheses. 
Signal shapes for the Higgs boson in decays involving 
the inclusive transition from \PGyP{2S} to \JPsi 
meson are modelled with a combination of the same functions as 
used for the Higgs boson directly decaying into ground 
state mesons (direct signal).
For the fits to the feed-down channels,
the background functions are identical and 
parameters are fixed to the ones from the  
previous direct signal fits.

Systematic uncertainties originate from imperfect 
knowledge of the detector and imperfect signal modeling. 
Systematic uncertainties considered in this analysis are 
listed below.

\renewcommand{\labelenumi}{\roman{enumi})}
\begin{enumerate}

\item The integrated luminosities for the 2016, 2017, and 2018 data-taking years have 1.2--2.5\% 
individual uncertainties~\cite{CMS:2021xjt,LUM-17-004,LUM-18-002}, 
while the overall uncertainty for the 2016--2018 period is 1.6\%. 

\item The differences between data and simulation for the trigger,
offline muon reconstruction, identification, and isolation efficiencies are corrected by
reweighting the simulated events with data-to-simulation correction factors,
which are obtained with the tag-and-probe method
using \JPsi (or \PZ)~$\to \MM$  events. 
The resulting scale
factors in muon identification, isolation, and trigger efficiencies are observed to deviate 
from unity by less than 2(2), 0.5(0.5) and 1(3)\% for the $\PZ\JPsi$ ($\mathrm{QQ}$) 
channel. Analogously, the uncertainty in the electron reconstruction, 
identification and trigger efficiency is found to be 
about 2\%~\cite{Khachatryan:2015hwa}.

\item The relative difference in the four-lepton vertex criterion
between data and simulation is evaluated with $\PZ\JPsi$ (\JPsi pair) 
event samples. It is found to be less than 2(3)\% for the $\PZ\JPsi$ 
($\mathrm{QQ}$) channel.

\item Differences in the lepton momentum scale and resolution
in data and simulation are estimated from \JPsi and \PZ dilepton
signals and extrapolated to the four-lepton signals.
The systematic uncertainty is estimated as the relative change in 
the upper limit when varying the signal mass mean and width by these
differences. They are found to be less than 1(3)\% in the $4\Pgm$ 
($2\Pe2\Pgm$) channel.

\item The theoretical uncertainties in the production cross section for the 
\PH (\PZ boson) are $\pm 3.2\%$ ($\pm 1.7\%$) due to the choice
of the PDF and the value of the strong coupling 
constant~\cite{LHC-YR4,Ball:2017nwa,Butterworth:2015oua},  
and $^{+4.6}_{-6.7}\%$ ($\pm 3.5\%$) due to the  
renormalization and factorization scale
choice~\cite{Butterworth:2015oua,Martin2009,PhysRevD.82.074024,BALL2011296}.

\item A common parameterization for each signal model is used for the entire run period. 
The relative uncertainty in the signal model due to the change in detector conditions in 
each year is determined to be 1(2)\% for the $\PZ\JPsi$ ($\mathrm{QQ}$) channels.

\item The background is alternatively parameterized
with a second order Chebyshev polynomial or a power law function.
The relative uncertainty due to the choice of the background function
is found to be negligible.

\item The uncertainties in the \PZ, \JPsi, and \PGU branching fractions to lepton pairs,
and $\mathcal{B}(\PGyP{2S}\to\JPsi + X)$ are taken from Ref.~\cite{pdg}. 

\end{enumerate}

\begin{table*}[htp!]
\centering
\topcaption{\label{results}Exclusion limits at 95\% \CL on the branching fractions of the
\PH and \PZ boson decays. The second column lists
the observed limits for the case that both intermediate particles are longitudinally polarized
($\lambda_\theta = -1$) as described in the text.
The third column shows the median 
expected limits with the upper and lower bounds in the expected 68\% \CL intervals.
The last two columns list observed upper limits for unpolarized ($\lambda_\theta = 0$) 
and transversely polarized ($\lambda_\theta = +1$) intermediate particles.}
\renewcommand{\arraystretch}{1.3}
\begin{tabular}{lccrr}
 Process & Observed & Expected & \multicolumn{2}{c}{Observed} \\
 \hline
Higgs boson channel         & Longitudinal & Longitudinal & Unpolarized & Transverse \\
$\mathrm{\mathcal{B}(\PH\to\PZ\JPsi)}$ & $1.9\times 10^{-3} $ & $(2.6^{+1.1}_{-0.7}) \times 10^{-3}$ & $2.4\times 10^{-3}$   & $2.8\times 10^{-3}$ \\
$\mathcal{B}(\PH\to\PZ\PGyP{2S})$ & $6.6\times 10^{-3} $ & $(7.1^{+2.8}_{-2.0}) \times 10^{-3}$ & $8.3\times 10^{-3}$   & $9.4\times 10^{-3}$  \\
$\mathrm{\mathcal{B}(\PH\to\JPsi\JPsi)}$ & $3.8\times 10^{-4} $ & $(4.6^{+2.0}_{-0.6}) \times 10^{-4}$  & $4.7\times 10^{-4}$    & $5.2\times 10^{-4}$  \\
$\mathcal{B}(\PH\to\PGyP{2S}\JPsi)$ & $2.1 \times 10^{-3} $ & $(1.4^{+0.6}_{-0.4}) \times 10^{-3}$ & $2.6\times 10^{-3}$  &  $2.9\times 10^{-3}$ \\
$\mathcal{B}(\PH\to\PGyP{2S}\PGyP{2S})$ & $3.0\times 10^{-3} $ & $(3.3^{+1.5}_{-0.9}) \times 10^{-3}$ &  $3.6\times 10^{-3}$  &  $4.7\times 10^{-3}$ \\
$\mathrm{\mathcal{B}(\PH\to\PGUP{nS}\PGUP{mS})}$  & $3.5\times 10^{-4} $ & $(3.6^{+0.2}_{-0.3}) \times 10^{-4}$ & $4.3\times 10^{-4}$  & $4.6\times 10^{-4}$  \\
$\mathrm{\mathcal{B}(\PH\to\PGUP{1S}\PGUP{1S})}$  & $1.7\times 10^{-3} $ & $(1.7^{+0.1}_{-0.1}) \times 10^{-3}$ &  $2.0\times 10^{-3}$  & $2.2\times 10^{-3}$  \\[\cmsTabSpace]
\PZ boson channel & & & &\\
$\mathrm{\mathcal{B}(\PZ\to\JPsi\JPsi)}$ & $11\times 10^{-7} $ & $(9.5^{+3.8}_{-2.6}) \times 10^{-7}$ & $ 14\times 10^{-7}$    & $ 16\times 10^{-7}$ \\
$\mathrm{\mathcal{B}(\PZ\to\PGUP{nS}\PGUP{mS})}$ & $3.9\times 10^{-7} $ & $(4.0^{+0.3}_{-0.3}) \times 10^{-7}$ & $ 4.9\times 10^{-7}$    & $ 5.6\times 10^{-7}$  \\
$\mathrm{\mathcal{B}(\PZ\to\PGUP{1S}\PGUP{1S})}$  & $1.8\times 10^{-6} $ & $(1.8^{+0.1}_{-0.0}) \times 10^{-6}$ & $2.2\times 10^{-6}$   & $2.4\times 10^{-6}$ \\
\end{tabular}
\end{table*}

\section{Results}
\label{sec:results}
No evidence for the Higgs or \PZ boson signal is found
in any of these channels.
The results of this analysis are 
presented as upper limits on the branching fractions and 
are set at the 95\% \CL. 
Limits are determined with the modified frequentist
$\CLs$ criterion, in which the profile 
likelihood ratio modified for upper limits
is used as the test statistic~\cite{junk,read,combiner}. 
Systematic uncertainties are incorporated in the likelihood 
as nuisance parameters.
The observed and median expected exclusion limits for the
branching fractions at 95\% \CL for the 
\PH and \PZ boson decays are listed in Table~\ref{results}.
Figures~\ref{fig:kineplot1ZJ}--\ref{fig:kineplot1YY} 
show the distributions for simulated boson
signals in different search channels as dashed lines.
The signals are normalized to their observed 95\% upper limit
branching fractions.

The results for $\mathcal{B}(\PH\to\PZ\JPsi)$
and $\mathcal{B}(\PH\to\PZ\PGyP{2S})$ each are obtained
by combining the channels with $\PZ\to\EE$ and $\PZ\to\MM$.  
The values for $\mathcal{B}(\JPsi\to\MM)$, 
$\mathcal{B}(\PZ\to\Pell^{+}\Pell^{-})$ and 
$\mathcal{B}(\PGyP{2S}\to\JPsi)$ are taken 
from Ref.~\cite{pdg}. This analysis does not 
distinguish between the three \PGUP{nS} states.
To calculate their contribution to the corresponding 
\PH and \PZ boson branching fractions, the coupling 
strength of the bosons to any \PGUP{nS}
pairing is assumed to be the same. All \PGU 
states can directly decay into muon pairs with the 
different branching fractions taken from 
Ref.~\cite{pdg}. In addition, it is assumed 
that the \PGU states could be the result of a one step transition 
$\PGUP{3S}\to\PGUP{2S}$, $\PGUP{3S}\to\PGUP{1S}$, 
or $\PGUP{2S}\to\PGUP{1S}$ 
before decaying into muons~\cite{pdg}.
Consequently, in the $\PGUP{1S}\PGUP{1S}$ channel, the 
feed-down transitions from \PGUP{3S} and
\PGUP{2S} to \PGUP{1S} are included. 
The observed upper limit branching fractions at 95\% \CL agree with 
the expected limits. In the case of  $\PH\to\PZ\JPsi$, the upper limit
branching fraction is about 800 times higher than the SM prediction~\cite{Alte:2016yuw}.  
In the $\PH\to\PGUP{nS}\PGUP{mS}$ channel, the observed upper 
limit branching fraction is found to be about one order of magnitude 
higher than SM predictions that assume dominance of direct quark couplings~\cite{soni1979}.
The factors are larger for all other channels. Tabulated results are available in the HEPData record for this analysis~\cite{HEPData}.

\section{Summary}
This Letter presents the first search for decays of the
Higgs boson (\PH) into a \PZ  boson and a \JPsi meson 
in four-lepton final states.  
Data from proton-proton collisions at $\sqrt{s}=13\TeV$, corresponding
to an integrated luminosity of about 138~\fbinv, are used.
Using the same data, decays of the Higgs and \PZ boson into
quarkonium pairs are also searched for.
No excess of a Higgs or \PZ boson signal above background 
is found in any of the searched channels
and upper limits on branching fractions ($\mathcal{B}$) at the 95\% confidence
level for various polarization scenarios are set.
The Higgs boson decay is also searched for in channels
where, before decaying into muon pairs, one or both \JPsi mesons 
could be the result of an inclusive \PGyP{2S} to \JPsi transition, 
and the $\PGUP{nS} (\mathrm{n}=1,2)$ mesons could be the result of inclusive
transitions from $\PGUP{nS} (\mathrm{n}=2,3)$ mesons.
The observed upper limits for the Higgs and \PZ boson decays
for longitudinally polarized mesons are
$\mathcal{B}(\PH\to\PZ\JPsi) < 1.9\times 10^{-3}$,
$\mathcal{B}(\PH\to\PZ\PGyP{2S} )< 6.6\times 10^{-3}$,
$\mathcal{B}(\PH\to\JPsi\JPsi) < 3.8\times 10^{-4}$,
$\mathcal{B}(\PH\to\PGyP{2S}\JPsi) < 2.1 \times 10^{-3}$, 
$\mathcal{B}(\PH\to\PGyP{2S}\PGyP{2S}) < 3.0\times 10^{-3}$,
$\mathcal{B}(\PH\to\PGUP{nS}\PGUP{mS}) < 3.5\times 10^{-4}$,
$\mathcal{B}(\PH\to\PGUP{1S}\PGUP{1S}) < 1.7\times 10^{-3}$,
$\mathcal{B}(\PZ\to\JPsi\JPsi)$ $< 11\times 10^{-7}$, 
$\mathcal{B}(\PZ\to\PGUP{nS}\PGUP{mS})   < 3.9\times 10^{-7}$,
and $\mathcal{B}( \PZ\to\PGUP{1S}\PGUP{1S}) < 1.8\times 10^{-6}$.
The observed upper limit branching fraction for $\PH\to\PZ\JPsi$ is about 
800 times the value predicted by the standard model~\cite{Alte:2016yuw,2014PhLB_7,gao}.
For $\PH\to\PGUP{nS}\PGUP{mS}$ the upper limit is about 
one order of magnitude higher than predicted
by earlier standard model calculations~\cite{soni1979}.
  
\begin{acknowledgments}
We congratulate our colleagues in the CERN accelerator departments for the excellent performance of the LHC and thank the technical and administrative staffs at CERN and at other CMS institutes for their contributions to the success of the CMS effort. In addition, we gratefully acknowledge the computing centers and personnel of the Worldwide LHC Computing Grid and other centers for delivering so effectively the computing infrastructure essential to our analyses. Finally, we acknowledge the enduring support for the construction and operation of the LHC, the CMS detector, and the supporting computing infrastructure provided by the following funding agencies: BMBWF and FWF (Austria); FNRS and FWO (Belgium); CNPq, CAPES, FAPERJ, FAPERGS, and FAPESP (Brazil); MES and BNSF (Bulgaria); CERN; CAS, MoST, and NSFC (China); MINCIENCIAS (Colombia); MSES and CSF (Croatia); RIF (Cyprus); SENESCYT (Ecuador); MoER, ERC PUT and ERDF (Estonia); Academy of Finland, MEC, and HIP (Finland); CEA and CNRS/IN2P3 (France); BMBF, DFG, and HGF (Germany); GSRI (Greece); NKFIH (Hungary); DAE and DST (India); IPM (Iran); SFI (Ireland); INFN (Italy); MSIP and NRF (Republic of Korea); MES (Latvia); LAS (Lithuania); MOE and UM (Malaysia); BUAP, CINVESTAV, CONACYT, LNS, SEP, and UASLP-FAI (Mexico); MOS (Montenegro); MBIE (New Zealand); PAEC (Pakistan); MES and NSC (Poland); FCT (Portugal); MESTD (Serbia); MCIN/AEI and PCTI (Spain); MOSTR (Sri Lanka); Swiss Funding Agencies (Switzerland); MST (Taipei); MHESI and NSTDA (Thailand); TUBITAK and TENMAK (Turkey); NASU (Ukraine); STFC (United Kingdom); DOE and NSF (USA).
  
\hyphenation{Rachada-pisek} Individuals have received support from the Marie-Curie program and the European Research Council and Horizon 2020 Grant, contract Nos.\ 675440, 724704, 752730, 758316, 765710, 824093, 884104, and COST Action CA16108 (European Union); the Leventis Foundation; the Alfred P.\ Sloan Foundation; the Alexander von Humboldt Foundation; the Belgian Federal Science Policy Office; the Fonds pour la Formation \`a la Recherche dans l'Industrie et dans l'Agriculture (FRIA-Belgium); the Agentschap voor Innovatie door Wetenschap en Technologie (IWT-Belgium); the F.R.S.-FNRS and FWO (Belgium) under the ``Excellence of Science -- EOS" -- be.h project n.\ 30820817; the Beijing Municipal Science \& Technology Commission, No. Z191100007219010; the Ministry of Education, Youth and Sports (MEYS) of the Czech Republic; the Hellenic Foundation for Research and Innovation (HFRI), Project Number 2288 (Greece); the Deutsche Forschungsgemeinschaft (DFG), under Germany's Excellence Strategy -- EXC 2121 ``Quantum Universe" -- 390833306, and under project number 400140256 - GRK2497; the Hungarian Academy of Sciences, the New National Excellence Program - \'UNKP, the NKFIH research grants K 124845, K 124850, K 128713, K 128786, K 129058, K 131991, K 133046, K 138136, K 143460, K 143477, 2020-2.2.1-ED-2021-00181, and TKP2021-NKTA-64 (Hungary); the Council of Science and Industrial Research, India; the Latvian Council of Science; the Ministry of Education and Science, project no. 2022/WK/14, and the National Science Center, contracts Opus 2021/41/B/ST2/01369 and 2021/43/B/ST2/01552 (Poland); the Funda\c{c}\~ao para a Ci\^encia e a Tecnologia, grant CEECIND/01334/2018 (Portugal); the National Priorities Research Program by Qatar National Research Fund; MCIN/AEI/10.13039/501100011033, ERDF ``a way of making Europe", and the Programa Estatal de Fomento de la Investigaci{\'o}n Cient{\'i}fica y T{\'e}cnica de Excelencia Mar\'{\i}a de Maeztu, grant MDM-2017-0765 and Programa Severo Ochoa del Principado de Asturias (Spain); the Chulalongkorn Academic into Its 2nd Century Project Advancement Project, and the National Science, Research and Innovation Fund via the Program Management Unit for Human Resources \& Institutional Development, Research and Innovation, grant B05F650021 (Thailand); the Kavli Foundation; the Nvidia Corporation; the SuperMicro Corporation; the Welch Foundation, contract C-1845; and the Weston Havens Foundation (USA).
\end{acknowledgments}

\bibliography{auto_generated}

\cleardoublepage \appendix\section{The CMS Collaboration \label{app:collab}}\begin{sloppypar}\hyphenpenalty=5000\widowpenalty=500\clubpenalty=5000
\cmsinstitute{Yerevan Physics Institute, Yerevan, Armenia}
{\tolerance=6000
A.~Tumasyan\cmsAuthorMark{1}\cmsorcid{0009-0000-0684-6742}
\par}
\cmsinstitute{Institut f\"{u}r Hochenergiephysik, Vienna, Austria}
{\tolerance=6000
W.~Adam\cmsorcid{0000-0001-9099-4341}, J.W.~Andrejkovic, T.~Bergauer\cmsorcid{0000-0002-5786-0293}, S.~Chatterjee\cmsorcid{0000-0003-2660-0349}, K.~Damanakis\cmsorcid{0000-0001-5389-2872}, M.~Dragicevic\cmsorcid{0000-0003-1967-6783}, A.~Escalante~Del~Valle\cmsorcid{0000-0002-9702-6359}, R.~Fr\"{u}hwirth\cmsAuthorMark{2}\cmsorcid{0000-0002-0054-3369}, M.~Jeitler\cmsAuthorMark{2}\cmsorcid{0000-0002-5141-9560}, N.~Krammer\cmsorcid{0000-0002-0548-0985}, L.~Lechner\cmsorcid{0000-0002-3065-1141}, D.~Liko\cmsorcid{0000-0002-3380-473X}, I.~Mikulec\cmsorcid{0000-0003-0385-2746}, P.~Paulitsch, F.M.~Pitters, J.~Schieck\cmsAuthorMark{2}\cmsorcid{0000-0002-1058-8093}, R.~Sch\"{o}fbeck\cmsorcid{0000-0002-2332-8784}, D.~Schwarz\cmsorcid{0000-0002-3821-7331}, S.~Templ\cmsorcid{0000-0003-3137-5692}, W.~Waltenberger\cmsorcid{0000-0002-6215-7228}, C.-E.~Wulz\cmsAuthorMark{2}\cmsorcid{0000-0001-9226-5812}
% ...
\par}
\vskip\cmsinstskip
\dag:~Deceased\\
$^{1}$Also at Yerevan State University, Yerevan, Armenia\\
$^{2}$Also at TU Wien, Vienna, Austria\\
% ...
\end{sloppypar}
\end{document}